\documentclass[11pt,a4paper]{article}

\usepackage{jcappub}

\usepackage{amsmath, amssymb, amsthm, graphicx, epsfig, fancyhdr,epsfig, slashed, mathrsfs}

\usepackage{tikzsymbols}
\usepackage{natbib}
\usepackage{float}

\usepackage{mathtools} 

\usepackage{tikz,xcolor,hyperref}

\definecolor{lime}{HTML}{A6CE39}
\DeclareRobustCommand{\orcidicon}{
	\begin{tikzpicture}
	\draw[lime, fill=lime] (0,0) 
	circle [radius=0.2] 
	node[white] {{\fontfamily{qag}\selectfont \tiny ID}};
	\draw[white, fill=white] (-0.0625,0.095) 
	circle [radius=0.007];
	\end{tikzpicture}
	\hspace{-2mm}
}

\foreach \x in {A, ..., Z}{\expandafter\xdef\csname orcid\x\endcsname{\noexpand\href{https://orcid.org/\csname orcidauthor\x\endcsname}
			{\noexpand\orcidicon}}
}

\newcommand{\be}{\begin{equation}}
\newcommand{\ee}{\end{equation}}
\newcommand{\bea}{\begin{eqnarray}}
\newcommand{\eea}{\end{eqnarray}}

\usepackage{orcidlink}

\begin{document}

\title{Leptogenesis in Brane-modified cosmology: Signatures in primordial gravitational waves}

\author[a]{Amit Dutta Banik\,\orcidlink{0000-0003-2602-3529},}
\emailAdd{amitdbanik@gmail.com}
\author[b]{Arnab Paul\,\orcidlink{0000-0003-3498-6755}}
\emailAdd{arnab.paul@physics.iitm.ac.in}

\affiliation[a]{Physics and Applied Mathematics Unit, Indian Statistical Institute, 203 B.T. Road, Kolkata 700108, India}
\affiliation[b]{Centre for Strings, Gravitation and Cosmology,
Department of Physics, Indian Institute of Technology Madras, 
Chennai~600036, India}

\abstract{We investigate the implications of brane-inspired modifications of the evolutionary history of the early universe on the process of baryogenesis via leptogenesis. A modified cosmic history alters the evolution of the Boltzmann equations governing lepton asymmetry, providing the possibility of successful leptogenesis in regions of the parameter space inaccessible in the standard cosmological scenario. Furthermore, a modified cosmic history also alters the shape of an otherwise scale-invariant spectral energy density (SED) of primary gravitational waves (PGWs) produced during inflation. This establishes a novel yet indirect probe of high scale leptogenesis scenarios through the observation of the SED of PGWs via future observations. We consider two scenarios in this work with single and multiple epochs of stiff equation of state(s) after inflation and before the epoch of radiation domination. We identify the parameter space where successful leptogenesis is possible, along with the possible observability of the PGWs, hence providing an indirect window into this leptogenesis scenario through PGWs.}

\maketitle

\section{Introduction}

Our theoretical and observational understanding of the thermal history of the universe dictates that during Big-Bang Nucleosynthesis (BBN) (at the temperature of $\sim1~$MeV), the energy density of the universe was dominated by radiation. However, the homogeneity of the universe at large scales and the isotropy of the observed cosmic microwave background (CMB) temperature map (with tiny anisotropies) lead us to hypothesize that much earlier than BBN, the universe underwent an accelerated expansion. This accelerated expansion, namely inflation, is often driven by the energy density of a slowly-rolling scalar field, namely the inflaton. Since the fundamental BB-mode polarization patterns in the CMB are yet to be detected, we have limits on the maximal Hubble rate during inflation $H_I< 5 \times 10^{13}$~GeV, or equivalently the maximal temperature of the universe of about $\sim10^{16}$~GeV \cite{Ade:2018gkx, Akrami:2018odb,BICEP:2021xfz}. 

However, the known evolutionary history after the BBN and the assumption of a specific inflationary model do not completely fix the evolution of the early universe; the evolution history of the universe between the end of inflation and the onset of BBN \cite{Allahverdi:2020bys} remains unconstrained. 
In the standard cosmological scenario, the universe is dominated by radiation just after the end of inflation, described by an equation of state (e.o.s) of $w = 1/3$. However, the dynamics of the inflaton field at the end of inflation, or an epoch dominated by some other field(s) can trigger a {\it stiff}  e.o.s $w > 1/3$, during which the energy density of the universe redshifts faster than radiation. For example, if the universe is dominated by the kinetic energy of a fast-rolling scalar field, the e.o.s during this epoch is $w = 1$ \cite{Spokoiny:1993kt,Joyce:1996cp,Ferreira:1997hj,Joyce:thesis}.

The possibility that the universe has a {\it stiff}  e.o.s with $w > 1/3$ is particularly relevant for the observation of a stochastic background of gravitational waves of primordial origin. The possible sources of gravitational waves (GW) in the early universe are inflation, (p)reheating as well as first-order phase transitions, cosmic strings \cite{Caprini:2018mtu} etc., if present. The observation of such primordial GWs in the interferometers or through the pulsar-timing arrays would not only offer a unique probe of the high-energy particle physics models responsible for non-standard sources of GW, but also of the cosmological history encoded in the GW spectra through the e.o.s of the early universe. In this work, we consider GWs produced from quantum fluctuations during inflation only, hereafter mentioned as primary gravitational waves (PGWs) and ignore other possible sources. We also assume a simple inflationary scenario giving rise to a scale-invariant spectral energy density (SED) of PGWs across the modes that reenter the Hubble radius during radiation domination epoch.

The interesting aspect of an era with a {\it stiff} e.o.s is that it leads to a rise in the SED of PGWs in an otherwise flat primary inflationary GW spectrum. For a de Sitter inflationary scenario, the SED at the end of inflation behaves as $k^2$ and $k^4$ with wave-number $k$ for super-Hubble ($k<k_{\rm e}$) and sub-Hubble modes ($k>k_{\rm e}$), respectively, where $k_{\rm e}=a(\eta_{\rm e})H(\eta_{\rm e})$ is the mode that exits the Hubble radius at the end of inflation (subscript ``${\rm e}$" denotes end of inflation, $\eta$, $a$ and $H$ denote conformal time, scale factor and Hubble rate of the universe). The SED must be regularized \cite{Birrell:1982ix,
Parker:2009uva,Wang:2015zfa,Pi:2024kpw,Hoory:2025qgm,Hoory:2026clf} to cut off this sub-Hubble ($k>k_{\rm e}$) quartic rise and hence achieve a finite GW energy density. Subsequent evolution of the scale factor modifies the SED of the super-Hubble modes at the end of inflation. The SED calculated at any specific time $\eta_f$ always retains its quadratic structure at the super-Hubble scales ($k_f<a(\eta_f)H(\eta_f)$). The shape of the spectrum between $k_f$ and $k_{\rm e}$ obtains modulations depending on the evolution of scale factor between $\eta_{\rm e}$ and $\eta_f$. For modes entering during matter, radiation, or kination dominated epochs, the SED behaves  as $k^{-2}$, $k^{0}$ and $k$, respectively. Thus, a kination domination epoch between inflation and the start of radiation domination gives rise to a SED that has a linear rise at very small scales just below $k_{\rm e}$, and hence may be observable in the GW observations.
Extensive studies have been performed on the impact of kination that occurs right after inflation on PGWs~\cite{ Giovannini:1998bp, Giovannini:2009kg, Riazuelo:2000fc, Sahni:2001qp, Seto:2003kc, Tashiro:2003qp, Nakayama:2008ip, Nakayama:2008wy, Durrer:2011bi, Kuroyanagi:2011fy, Kuroyanagi:2018csn, Jinno:2012xb, Lasky:2015lej, Li:2016mmc, Saikawa:2018rcs, Caldwell:2018giq, Bernal:2019lpc, Figueroa:2019paj, DEramo:2019tit,Haque:2021dha,Li:2013nal, Li:2021htg}. 
 
Leptogenesis \cite{Fukugita:1986hr, Buchmuller:2004nz, Anisimov:2007mw, Davidson:2008bu, Buchmuller:2005eh, Baek:2013qwa, Davoudiasl:2015jja, Hernandez:2016kel, Guo:2016ixx}, is a mechanism for generation of matter-antimatter asymmetry in the universe via out of equilibrium decay of particles that violate CP asymmetry and lepton number conservation based on Shakharov conditions. Study of leptogenesis with non-standard cosmological evolution of the universe has been explored extensively in the literature \cite{Dutta:2018zkg,Chen:2019etb}. In this work, we will study the effect of non-standard e.o.s during the early universe on leptogenesis and PGW SED and hence study the possibility to indirectly probe high scale leptogenesis through observability of PGWs, which otherwise is difficult to probe. 
  
The paper is organized as follows. In section\,\ref{cosmo}, we discuss how scalar fields motivated by super-string theory change the cosmic evolution in the early universe and study its effect on leptogenesis and the PGW SED. Then, in section\,\ref{obser} we study the possibility of the modulated SED of PGWs to be observed in future GW observations. Then in section\,\ref{disc} we discuss the results and conclude.

\section{Scalar domination in early universe}
\label{cosmo}
In this section, we will discuss how the evolution history of the universe changes in the presence of additional scalar fields that dominate between inflation and BBN. In standard scenario, the energy density stored in the inflaton decays into radiation through reheating, and the energy density of the universe in this radiation-dominated (RD) epoch is given by,
\be
\rho_{\rm{RD}}=\frac{\pi^2}{30}g_*(T)T^4,  
\ee
where $g_*(T)$ is the effective relativistic degrees of freedom at temperature $T$. The Hubble rate $H_{\rm{RD}}$ during this epoch is then given by,
\be
H_{\rm{RD}}=\sqrt{\frac{8\pi}{3M_{\rm P}^2}\rho_{\rm{RD}}}=1.66 \sqrt{g_*} \frac{T^2}{M_{\rm P}}\, ,
\label{Hub}
\ee
with $M_{\rm P}=1/G=1.22\times 10^{19}$ GeV, where $G$ is the Newton's constant.
Now, instead of this standard cosmological scenario, we consider the case of non-standard (NS) expansion (encoded in e.o.s $w$) of the universe. For such an epoch, the energy density evolves as,
\be
\rho_{\rm{NS}}\propto a^{-3(1+w)}\, ,
\label{rhophi}
\ee 
where $a$ is the scale factor of the universe. Using comoving entropy conservation, i.e. $sa^3$ is conserved (where $s=\frac{2\pi^2}{45}g_{s}T^3$) and assuming at $T=T_{\rm R}$, 
$\rho_{\rm {NS}}=\rho_{\rm{RD}}$, one can write
\be
\rho_{\rm{NS}}(T)=\rho_{\rm{RD}}(T_{\rm R})\bigg(\frac{g_{s}(T)}{g_{s}(T_{\rm R})}\bigg)^{1+w}\bigg(\frac{T}{T_{\rm R}}\bigg)^{3w+3}\,\, .
\label{phirad1}
\ee
Therefore, the Hubble rate $H_{\rm{NS}}\propto\sqrt{\rho_{\rm{RD}}+\rho_{\rm{NS}}}$, in non-standard expansion era is given by,
\be
\label{hnew1}
H_{\rm{NS}}=1.66 \sqrt{g_*} \frac{T^2}{M_P}\bigg[1 +\bigg(\frac{T}{T_{\rm R}}\bigg)^{3w-1}\bigg]^{1/2}
=H_{\rm{RD}}\bigg[1+\bigg(\frac{T}{T_{\rm R}}\bigg)^{3w-1}\bigg]^{1/2}\,\, ,\\
\ee
where, for simplicity, we have assumed that $g_*$ remains the same and $g_*\simeq g_{s}$ during this early epoch. This expression captures the fact that, for $T\gg T_{\rm R}$, \textit{i.e.}, during the kination-dominated epoch with $w=1$, the expansion rate of the universe with respect to temperature is faster than standard radiation-dominated era. The lower bound on $T_{\rm R}$ comes from the fact that the universe should become radiation-dominated before the BBN epoch, i.e., before the temperature reaches $T_{\rm{BBN}}$. Since we are working in the context of leptogenesis at high temperature, we conservatively assume $T_{\rm R}\gg T_{\rm{BBN}}$. 
An epoch dominated by a kination-like energy density ($w=1$), as shown in Eq.~(\ref{rhophi}), can be achieved with a scalar field $\phi$ minimally coupled to gravity \cite{DEramo:2017gpl}. It is also possible to consider a scenario where energy density of multiple scalar fields dominate sequentially before the start of radiation domination era.
In such a scenario with two scalar fields, the energy density of scalar fields follows the sequence
\begin{eqnarray}
\label{seq}
    \rho_{\phi_2}>\rho_{\phi_{1}} \mbox{ for } T>T_2 ,\\
    \rho_{\phi_2}<\rho_{\phi_{1}} \mbox{ for } T<T_2 .
\end{eqnarray}
With comoving entropy being conserved, for two such scalar fields, the Hubble parameter can be expressed as
\begin{equation}\label{rho_new} 
    \rho_{\rm{NS}}(T) = 
    \rho_{\rm{RD}}\left\{\left(\frac{T}{T_{\rm R}}\right)^{3w_1-1}\left[1+\left(\frac{T}{T_2}\right)^{3(w_2-w_1)}\right]\right\},
\end{equation}
where $w_2>w_1$, which ensures that the energy density is governed by the $\phi_2$ scalar field when $T\ge T_2$. Similarly, in the range $T_2>T>T_{\rm R}$, $\phi_1$ dominates the energy density of the universe until the radiation energy density dominates the scalar fields at $T=T_{\rm R}$. Therefore, the modified Hubble parameter is now expressed as
\begin{equation}\label{Hnew}   
    H_{\rm{NS}}=H_{\rm{RD}}\left\{1+\left(\frac{T}{T_{\rm R}}\right)^{3w_1-1}\left[1+\left(\frac{T}{T_2}\right)^{3(w_2-w_1)}\right]\right\}^{1/2}\,
\end{equation}

We now briefly describe the origin of these scalar fields driving the kination-like epoch after the end of inflation. These scalar fields (decoupled from other fields) may originate from super-string motivated orientifold models related to D-branes when the higher spatial dimensions are reduced into four dimensions~\cite{DiMarco:2018bnw,Maharana:2017fui,Koivisto:2013fta}.
Scalar fields in this scenario basically represent the degrees of freedom arising from the dynamical fluctuations of the D-branes in the transverse directions. These scalar fields that interact with gravity only at the tree level, are referred to as moduli and get stabilized after the partial breaking of the symmetry. Energy density of these scalar fields are dominated by kinetic terms, hence giving rise to kination-like e.o.s in the post inflationary phase. The energy densities follow
\begin{equation}
    {\rho_{\phi_i}}+3H\rho_{\phi_i}(1+w_i)=0\, .
\end{equation} 
These scalar fields are assumed not only to be disconnected from the longitudinal degrees of freedom of D-branes, but also to have no interactions with visible and dark sector or mutual interactions, as previously mentioned. However, coupling to the inflaton is possible.
The dynamics of these fields give rise to different phases of kination with $w_i>1/3$. As mentioned before, scalar fields with $w_i>1/3$ dilute faster than radiation as the temperature of the universe decreases. This ensures these unwanted relics become negligible eventually and thus avoids conflicts with BBN.

\begin{figure}[H]
\begin{center}
\includegraphics[width=0.49\textwidth]{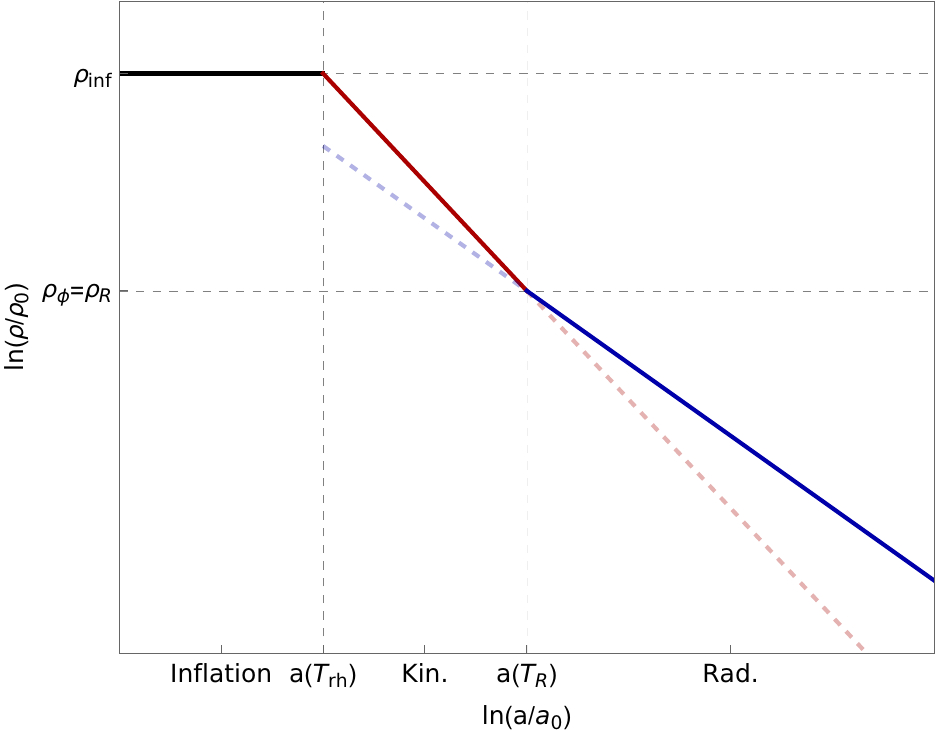}
\includegraphics[width=0.49\textwidth]{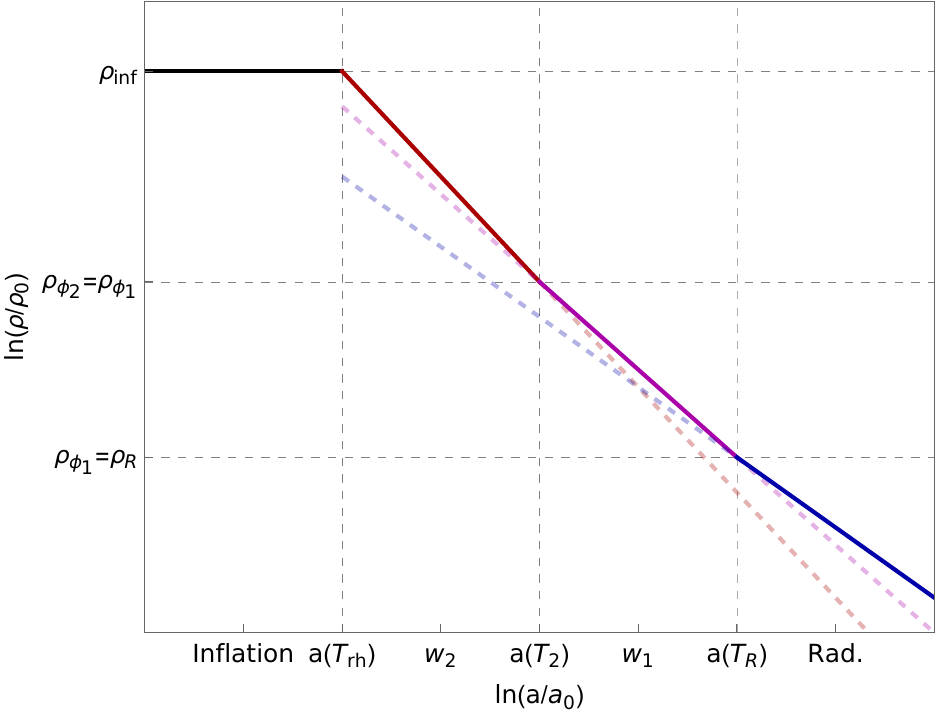}
	\caption{\it In these plots, we have shown the schematic illustrations of the evolution of energy density of the different components of the universe for the two scenarios discussed in this work. The black line corresponds to inflaton energy density, whereas the blue line corresponds to the energy density of the radiation bath. In between these two epochs, in the left panel we have an epoch when a scalar field dominates with $w=1$, shown with the red line. In the right panel, we show the energy density of two scalars dominating one after another with e.o.s $w_2=1$ and $w_1=2/3$ respectively, shown with red and magenta lines. $\rho_0$ and $a_0$ correspond to some scaling of energy density and scale factor, respectively.}
\label{fig:rho}
\end{center}
\end{figure}

We impose some conditions on the reheating temperature $T_{\rm rh}$\footnote{We define the reheating temperature $T_{\rm rh}$ to be the temperature of the radiation soup at the end of inflation, i.e., the maximum temperature attained by the radiation bath. It is much higher than the temperature $T_{\rm R}$ at which the radiation energy density dominates over the scalar fields eventually.} based on the maximal energy density stored in a scalar field. We assume that the energy density of a scalar field does not exceed some scale $M^4<M_{\rm P}^4$. This imposes an upper bound on the energy density of a scalar field at $T_{\rm rh}$, $\rho_{\phi}(T_{\rm rh})\leq M^4$ for the scenario with one scalar. One can easily obtain an upper limit on $T_{\rm rh}$, given by, 
\begin{equation}\label{eqn: bound_1}
    T_{\rm rh} \leq\left(\frac{30}{\pi^2 g_*(T_{\rm R})}\right)^{\frac{1}{3(1+w)}} M \left(\frac{T_{\rm R}}{M} \right)^{\frac{3{w}-1}{3(1+{w})}}\, ,
\end{equation}
where $T_{\rm R}$ is the temperature of transition at which the energy density of the field becomes the same as that of radiation. 
In a similar fashion, as described earlier, for two such scalar fields giving rise to kination-like epochs, the upper limit on $T_{\rm rh}$ translates into
\begin{equation}\label{enq: bound_2}
    T_{\rm rh} \leq \left(\frac{30}{\pi^2 g_*(T_{R})}\right)^{\frac{1}{3(1+w_2)}} M \left( \frac{T_{\rm R}^{3w_1-1} T_2^{3(w_2-w_1)}}{M^{3w_2-1}} \right)^{\frac{1}{3(1+w_2)}}\, .
\end{equation}
where, $w_2>w_1$ and $T_{\rm rh}>T_2>T_{\rm R}$. To clarify the notation, in Fig.~\ref{fig:rho} we have shown the schematic illustrations of the two scenarios discussed in this work.

\subsection{Leptogenesis in Brane cosmology}
\label{LEP}

In this section, we discuss the effect of kination-like epochs in the Boltzmann equations of the standard leptogenesis process. We consider the usual unflavored leptogenesis via right handed neutrinos (RHN) with the type-I seesaw mechanism (with heavy RHN mass $\simeq 10^9-10^{11}$ GeV). 
The Type-I seesaw mechanism for small neutrino mass is invoked by the insertion of heavy Majorana neutrinos with hypercharge zero, which allows lepton number violation in a beyond Standard Model (SM) theory with the following interaction Lagrangian,
\be
{\cal L}=-Y_{ij}\bar{L_i}\tilde{H} N_{j} - \frac{1}{2}M_j\bar{N^c}_jN_j + h.c.\,\, ,
\label{L}
\ee
where $H$ is the Standard Model Higgs doublet, $L$ is a SM lepton doublet, $Y_{ij}$ denotes the coupling between heavy RHN ($N$) and SM leptons through the Higgs doublet, and the masses of RHNs are given by $M_j$. For simplicity, we consider the mass matrix for heavy Majorana neutrinos to be diagonal and hierarchical $M_3>M_2>M_1$. The light neutrino mass is obtained from the type-I seesaw mechanism $M_{\nu}=-m_{D}^T{\cal M}^{-1}m_{D}$, with the Dirac neutrino mass matrix $m_D$ and the diagonal mass matrix ${\cal M}$ of the heavy RHNs. The decay of heavy RHNs, with the interference of tree and loop-level processes, gives rise to CP asymmetry $\varepsilon$ for complex Yukawa coupling and is given by,

\begin{align}
\varepsilon & =-\frac{3}{16\pi}\frac{1}{(Y^{\dagger}Y)_{11}}\sum_{j=2,3}{\rm Im}[(Y^{\dagger}Y)^2_{1j}]\frac{M_1}{M_j}\,\, . 
\label{CPasy}
\end{align}
considering $M_3>M_2>M_1$. The CP asymmetry $\varepsilon$ expressed in Eq.~(\ref{CPasy}) can be parameterized by Casas-Ibarra formalism~\cite{Casas:2001sr}
\be
\vert\varepsilon\vert< \frac{3}{8\pi v^2}M_1m_{\nu}^{\rm max}\,\, ,
\label{limit}
\ee
where $m_{\nu}^{\rm max}\simeq 0.1$ eV is the largest active neutrino mass \cite{ParticleDataGroup:2018ovx} and $v=246$ GeV, is the vacuum expectation value of SM Higgs. The typical RHN mass obtained from Eq.~(\ref{limit}) is found to be $\mathcal{O}(10^{10}~ \rm GeV)$ for the choice of $\varepsilon\simeq 10^{-6}$, in order to generate the baryon asymmetry in standard leptogenesis, and remain consistent with the neutrino mass limits. 

With the above set-up for vanilla leptogenesis, we now derive the expressions for the Boltzmann equations for leptogenesis from the lightest RHN $N_1$ of mass $M_1$. We begin with the Boltzmann equations for standard radiation-dominated leptogenesis,

\begin{eqnarray}
\frac{d Y_{N_1}}{dz}&=& 
-z\frac{\Gamma_1}{H_1}
\frac{K_1(z)}{K_2(z)}\left(Y_{N_1}-Y_{N_1}^{\rm eq}\right)\,,
\nonumber \\
\frac{d Y_{L}}{dz} &=&  - \frac{\Gamma_1}{H_1} 
\left [\varepsilon  z \frac{K_1(z)}{K_2(z)}(Y_{N_1}^{\rm eq}- Y_{N_1})  +  
\frac{z^{3} K_1(z)}{4} Y_L  \right]\, \, .
\label{lep_stand}
\end{eqnarray}
In the above expression, $Y_i=n_i/s$ is the ratio between number density and entropy density, with their equilibrium abundance represented by $Y_{i}^{\rm eq}$, $H_1=H_{\rm{RD}}(T=M_1)$, $z=M_1/T$, $\Gamma_1=\frac{M_1}{8\pi}(Y^{\dagger}Y)_{11}$ is the decay width of $M_1$ and $K_i$ with $i=1,2$ are modified Bessel functions.
Using Eq.~(\ref{limit}) and $m_{\nu}^{\rm max}\simeq 0.1$ eV, one obtains a simple relation $\varepsilon \simeq 10^{-16} M_1/\rm GeV$. Considering a case of two massive RHNs, the asymmetry parameter simplifies to 
\begin{equation}
\varepsilon =-\frac{3}{16\pi}\frac{1}{(Y^{\dagger}Y)_{11}}{\rm Im}[(Y^{\dagger}Y)^2_{12}]\frac{M_1}{M_2}.
\end{equation}
Assuming the Yukawa couplings $Y_{ij}$ of Eq.~(\ref{L}) are of the same order (i.e., RHNs couple with leptons with equal strength,  resulting in Im$[(Y^{\dagger}Y)_{12}]\simeq (Y^{\dagger}Y)_{11}$), one obtains $(Y^{\dagger}Y)_{11}\cong (16\pi/3)\varepsilon M_2/M_1$. The effective washout factor (known as the decay parameter) is found to be in the strong washout regime 
\begin{equation}
K=\frac{\Gamma_1}{H_1} \simeq \frac{1}{8\pi}\frac{(Y^{\dagger}Y)_{11}}{1.66g_*^{1/2}}\frac{M_{\rm P}}{M_1}\equiv\frac{2}{3}\frac{M_{\rm P}M_2}{1.66g_*^{1/2}M_1}\frac{10^{-16}}{\rm GeV}\leq 600
\label{decpar}
\end{equation}
for $M_1:M_2=1:10$. Interestingly, the value of the decay parameter becomes independent of $M_1$, and remains unaltered once the ratio $M_1:M_2$ is fixed. Solutions to Eq.~(\ref{lep_stand}) above, with initial equilibrium RHN abundance ($Y_{N_1}^{\rm{in}}=Y_{N_1}^{\rm eq}$), provide the lepton asymmetry yield at present-day ($T=T_{0}$) in standard cosmological evolution without a kination-like epoch. In standard cosmological evolution, although one can generate the correct order of baryon asymmetry with heavy RHNs ($M_1\simeq 10^{9}-10^{11}$ GeV) initially, while being consistent with neutrino mass, we will show that the high washout effect due to large $K$ depletes the asymmetry significantly.  

\subsubsection{Single field}

The above issue may be alleviated when a kination-like epoch is introduced. A kination-like epoch modifies the Hubble rate of Eq.~(\ref{Hub}) with a modified expression given by Eq.~(\ref{hnew1}) or Eq.~(\ref{Hnew}), depending on the number of active scalar fields.
Applying the modifications to the Hubble parameter, with the influence of a single scalar field, the Boltzmann equations (BEs) for leptogenesis are expressed as,

\begin{eqnarray}
\frac{d Y_{N_1}}{dz}&=& 
-z\frac{\Gamma_1}{H_1}\frac{1}{J_1}
\frac{K_1(z)}{K_2(z)}\left(Y_{N_1}-Y_{N_1}^{\rm eq}\right)\,,
\nonumber \\
\frac{d Y_{L}}{dz} &=&  - \frac{\Gamma_1}{H_1} \frac{1}{J_1}
\left (\varepsilon  z \frac{K_1(z)}{K_2(z)}(Y_{N_1}^{eq}- Y_{N_1})  +  
\frac{z^{3} K_1(z)}{4} Y_L  \right)\, \, .
\label{asy}
\end{eqnarray}
where $J_1$ is written as,

\begin{equation}\label{factor}
    {J_1}=\left\{1+\left(\frac{M_1}{T_{\rm R} z}\right)^{3w-1}\right\}^{1/2}\, .
\end{equation}

Since $J_1\gg1$, it reduces the washout effect as $\Gamma_1/(H_1 J_1)\approx 600/J_1$ and increases the net lepton asymmetry, assuming simple unflavored leptogenesis is valid~\footnote{With the scalar fields in effect, the faster expansion delays the equilibration of charged lepton Yukawa interactions. For example, $\tau$ lepton interaction becomes significant at $T_\tau\simeq 10^{11}$ GeV in standard radiation-dominated expansion of the universe as it reaches $\Gamma_\tau/H_{\rm RD}\geq 1,~\Gamma_\tau\approx5\times 10^{-3}y_\tau^2 T$ with $y_\tau$ being $\tau$ lepton Yukawa coupling. However, in our case with a single scalar field, this interaction becomes relevant at $T_\tau\leq 2.15\times 10^7 T_{\rm R}^{1/3}$ for $w=2/3$ as the expansion rate becomes $H_{\rm NS}=H_{\rm RD}(1+T/T_{\rm R})^{1/2}$. Therefore, for $T_{\rm R}\simeq 10^8$ GeV, the $\tau$ lepton interaction becomes significant at $T_\tau\simeq 10^{10}$ GeV, allowing vanilla leptogenesis for $M_1\geq10^{11}$ GeV. Whereas for $T_{\rm R}\sim 10^5$ GeV the $\tau$ lepton interaction enters equilibrium at $T_\tau=10^9$ GeV. Similarly, for $w=1$, one finds $T_\tau\simeq 3.16\times 10^9$ GeV for $T_{\rm R}=10^8$ GeV, further broadening the regime of unflavored leptogenesis. As a result, the scope of vanilla leptogenesis increases even for lower masses of $M_1$. Eventually, this is also valid when the effect of multiple fields is taken into account. Therefore, throughout the work, we consider unflavored leptogenesis only in our analysis.}.
\subsubsection{Multiple fields}
In a similar fashion, in the presence of two scalar fields where $\phi_2$ dominates the energy density above temperature $T>T_2$ and $\phi_1$ dominates energy density within the temperature range $T_2>T>T_{\rm R}$, the Boltzmann equations governing leptogenesis are modified as

\begin{eqnarray}
\frac{d Y_{N_1}}{dz}&=& 
-z\frac{\Gamma_1}{H_1}\frac{1}{J_2}
\frac{K_1(z)}{K_2(z)}\left(Y_{N_1}-Y_{N_1}^{\rm eq}\right)\,,
\nonumber \\
\frac{d Y_{L}}{dz} &=&  - \frac{\Gamma_1}{H_1} \frac{1}{J_2}
\left (\varepsilon  z \frac{K_1(z)}{K_2(z)}(Y_{N_1}^{eq}- Y_{N_1})  +  
\frac{z^{3} K_1(z)}{4} Y_L  \right)\, \, .
\label{asy2}
\end{eqnarray}
where $J_2$ is given as,

\begin{equation}\label{factor2}
    {J_2}=\left\{1+\left(\frac{M_1}{T_{\rm R} z}\right)^{3\omega_1}\left[1+\left(\frac{M_1}{T_{\rm R} x  z}\right)^{3(\omega_2-\omega_1)}\right]\right\}^{1/2}\, ,
\end{equation}
with $x=T_2/T_{\rm R}$. Similar to the single field scenario, for $J_2>J_1$, the effective washout parameter is considerably reduced. Therefore, multiple fields enhance the possibility of larger lepton asymmetry generation. Interestingly, $J_2$ carries the information of two new scales given by the ratios $x=T_2/T_{\rm R}$ and $T_{\rm R}/M_1$, which determine the length of the second kination-like epoch due to $\phi_1$ and the temperature at the end of $\phi_1$ domination scaled to the mass of the lightest right
handed neutrino, respectively. 

The lepton asymmetry $Y_L$ produced in leptogenesis is partially transferred into baryon asymmetry $Y_B$ via electroweak Sphalerons following~\cite{Davidson:2008bu},
\be  
Y_B=\frac{8n_f+4n_{\phi}}{22n_f+13n_{\phi}}Y_L\,\, ,
\label{relation}
\ee
where $n_{\phi}$ and $n_{f}$ denote the number of Higgs doublets and lepton doublets.
With three fermion generations and one Higgs doublet involved, Eq.~(\ref{relation}) becomes,
\be
Y_B=\frac{28}{79}Y_L\, .
\label{BL}
\ee
We require $Y_{B}\simeq(8-10)\times 10^{-11}$ \cite{ParticleDataGroup:2016lqr} to explain matter anti-matter asymmetry in the Universe. Extensive study of leptogenesis with similar scalar and multiple scalar fields has been done before in the literature \cite{Chen:2019etb,DiMarco:2022doy}. We focus on the observability aspect of high scale leptogenesis with gravitational waves, as the GW spectrum also gets modified in the presence of such kination-like epochs.

\subsection{Induced features in SED of PGWs}\label{sec:GWspec}

In this work, we consider GW produced only from quantum fluctuations of the metric during inflation, we do not take into account astrophysical sources or other primordial cosmological sources like (p)reheating or phase transitions. In standard models of inflation, the spectral energy density of the modes that reenter during radiation domination exhibits mostly scale-invariant behavior. 
In the simplest scenario, the universe enters a radiation-dominated phase just after inflation until the matter content surpasses the radiation. In this work, we are interested in scenarios where there is one or more epochs with an e.o.s $w\neq1/3$. These scenarios lead to a modulation in the otherwise scale-invariant  SED of PGWs. In particular, if the non-standard e.o.s $w>1/3$, i.e., the energy density falls faster than radiation, a blue tilt is imprinted in the SED. On the other hand, an epoch with $w<1/3$ would result in a red tilt in the SED. In this work, we will discuss two scenarios with one and two stiff epochs ($w>1/3$), as mentioned before. In this section, we discuss how the SED of PGWs is modified due to such non-standard evolutions.

\subsubsection{Single field}
We start with a scenario where the inflationary phase is followed by an epoch when the energy density of the universe is dominated by a scalar field, resulting in a stiff e.o.s. As the energy density of this scalar field falls faster than the radiation density, it is eventually overtaken by the radiation density. So, we end up with a scenario where the e.o.s of the universe changes from $w=-1$ during inflation to $w>1/3$ during kination-like epochs to eventually $w=1/3$ during radiation domination. The Hubble rate at the transition from kination to radiation domination is given by 
\begin{equation}
    H_{\rm R}^2=\frac{8\pi}{3 M_P^2}\frac{\pi^2}{30}g_{*,\rm R}T_{\rm R}^4,
\end{equation}
where $g_{*,\rm R}\equiv g_*(T_{\rm R})$ and $T_{\rm R}$ corresponds to the temperature of the thermal soup at the end of the kination-like epoch. The mode with wavelength equal to the Hubble radius at the end of kination is,
\begin{equation}
k_{\rm R} = H_{\rm R}\frac{a_{\rm R}}{a_{0}}  = H_{\rm R}\left(\frac{g_{s, \rm eq} }{g_{s,\rm R}} \right)^{\frac{1}{3}}\frac{T_{0}}{T_{\rm R}}  \,,
\label{eq:kke0}
\end{equation}
where $g_{s, \rm R}\equiv g_s(T_{\rm R})$ and $g_{s, \rm eq}\equiv g_s(T_{\rm eq}) = 2+2N_{\rm eff}(7/8)(4/11) = 3.938$, $T_{0} = 2.35\times 10^{-13}$ GeV denote the present-day temperature of photons. The second equality comes from the conservation of comoving entropy density.

As mentioned before, the initial power spectrum of the PGW is assumed to be scale-invariant in this work. For a scale-invariant initial power spectrum the SED of the modes that enter during radiation domination exhibits scale-invariant behavior. For modes that enter during a kination-like epoch, we get a blue-tilt \cite{Opferkuch:2019zbd,Haque:2021dha}. For such a scenario, the SED of PGW today is given by
\begin{equation}
	\Omega^{0}_{\rm GW} (k) = \Omega_{\rm GW}^{\rm 0, flat} \begin{cases}
		 1\,, & k < k_\text{R} \\ 
		\left(\frac{k}{k_\text{R}}\right)^{\frac{2(3w-1)}{1+3w}}\,, & k_\text{R} \leq k \leq k_{\rm e}\ \\
		0 &  k_{\rm e}<k
			\end{cases} \,.
	\label{eq:GWspecTot}
\end{equation}
The amplitude of the scale-invariant portion $\Omega_{\rm GW}^{\rm 0, flat}$ is given by,
\begin{equation}
\Omega_{\rm GW}^{\rm 0, flat} = \frac{\Omega_{\gamma}^{0}}{24}  \left(\frac{g_{s,\rm eq}}{g_{s,k}}\right)^{\frac{4}{3}} \left(\frac{g_{k}}{g_{\gamma}^{0}} \right) \frac{16}{\pi} \frac{H_{\rm e}^{2}}{M_P^{2}}\,,
\end{equation}
where $\Omega_{\gamma}^{0}$ is the fraction of energy in radiation today, $g_{\gamma}^{0} = 2$ and $g_{k}$ are the degrees of freedom when the $k$ mode re-entered the horizon. $H_{\rm e}$ corresponds to the Hubble parameter at the end of inflation and initiation of kination.

We would like to mention a few points here. Firstly, the modes that enter during matter domination, exhibit a red tilt, however, we are not interested in the SED corresponding to those large scales here. Another point is about the SED at very small scales, the modes that were always sub-Hubble. We have assumed that $\Omega^{0}_{\rm GW} (k)=0$ for those small modes ($k>k_{\rm e}$)\footnote{If regularization of SED is not considered, we get a unphysical quartic growth at small scales, i.e. $\Omega^{0}_{\rm GW} (k)\sim k^4$ for ($k>k_{\rm e}$). However, a properly regularized SED falls off exponentially with $k$ \cite{Pi:2024kpw}.}. The cutoff scale $k_{\rm e}$ of the SED is given by,
\begin{equation}
k_{\rm e} =H_{\rm e}\left(\frac{a_{\rm e}}{a_0}\right)=H_{\rm e}\left(\frac{a_{\rm R}}{a_0}\right)\left(\frac{a_{\rm e}}{a_{\rm R}}\right)= H_{\rm e} \left(\frac{g_{s, \rm eq} }{g_{s,\rm R}} \right)^{\frac{1}{3}} \left(\frac{T_{0}}{T_{\rm R}} \right) \left(\frac{H_{\rm R}^{2}}{2H_{\rm e}^{2}} \right)^{\frac{1}{3(1+w)}} \,,
\end{equation}
The third equality comes from the conservation of comoving entropy density, which leads to the relation $a_{\rm e}/a_{\rm R}=[H_{\rm R}^{2}/(2H_{\rm e}^{2})]^{1/(3+3w)}$.
The frequency $f$ of GW is related to the wave-number $k$ via the relation $f=c\,k/(2\pi)$, where $c$ is the speed of light.

\subsubsection{Multiple fields}
We now consider a scenario where inflation is followed by two back-to-back epochs of stiff e.o.s before the eventual radiation domination epoch. The Hubble rate and the physical mode corresponding to the transition from kination to radiation domination have been discussed before. The mode with a wavelength of the horizon size at the transition between the two kination-like epochs is given by,
\begin{equation}
k_{2}= H_{2}\left(\frac{a_2}{a_{0}} \right) = H_{2}\left(\frac{g_{s, \rm eq} }{g_{s,2}} \right)^{\frac{1}{3}}\left(\frac{T_{0}}{T_2} \right) \,,
\label{eq:kkm2}
\end{equation}
where the suffix $``2"$ denotes the transition from the first stiff e.o.s $w_2$ to the second stiff e.o.s $w_1$. The Hubble rate during this transition is given by,
\begin{equation}
    H_{2}^2=\frac{16\pi}{3 M_P^2}\frac{\pi^2}{30}g_{\rm R}T_{\rm R}^4\left(\frac{T_{2}}{T_{\rm R}}\right)^{3(w_1+1)}.
\end{equation}
The SED of PGWs today for this scenario is given by
\begin{equation}
	\Omega^{0}_{\rm GW} (k) = \Omega_{\rm GW}^{\rm 0, flat} \begin{cases}
		 1\,, & k < k_\text{R} \\ 
		\left(\frac{k}{k_\text{R}}\right)^{\frac{2(3w_1-1)}{1+3w_1}}\,, & k_\text{R} \leq k \leq k_{2} \\
		\left(\frac{k_\text{2}}{k_\text{R}}\right)^{\frac{2(3w_1-1)}{1+3w_1}}\left(\frac{k}{k_\text{2}}\right)^{\frac{2(3w_2-1)}{1+3w_2}}\,,& k_\text{2} \leq k \leq k_{\rm e}  \\
		0 &  k_{\rm e}<k
			\end{cases} \,
	\label{eq:GWspecTot2}
\end{equation}
where $\Omega_{\rm GW}^{\rm 0, flat}$ has been defined before.
The mode corresponding to the cutoff of the SED is given by
\begin{equation}
k_{\rm e}=H_{\rm e}\frac{a_{\rm e}}{a_0}=H_{\rm e}\left(\frac{a_{\rm R}}{a_0}\right)\left(\frac{a_2}{a_{\rm R}}\right)\left(\frac{a_{\rm e}}{a_{2}}\right)
\end{equation}
where $a_{\rm R}/a_0=\left(g_{s, \rm eq} /g_{s,\rm R} \right)^{\frac{1}{3}} \left(T_{0}/T_{\rm R} \right)$, $a_{\rm 2}/a_{\rm R}=\left(g_{s, \rm R} /g_{s,2} \right)^{\frac{1}{3}} \left(T_{\rm R}/T_{\rm 2} \right)$ and the ratio $a_{\rm e}/a_{2}$ is given by,
\begin{equation}
\frac{a_{\rm e}}{a_{2}}=\left[\frac{8\pi^3g_{\rm R}T_{\rm R}^4\left(T_{2}/T_{\rm R}\right)^{3(w_1+1)}}{90M_P^2 H_{\rm e}^{2}} \right]^{\frac{1}{3(1+w_2)}}.
\end{equation}

The total energy density stored in the PGW today ($\Omega_{\rm GW}^{0}$) is given by the integral of the SED, 
\begin{equation}
\Omega_{\rm GW}^{0} =  \int_{f_{\rm low}} ^{f_{\rm e}} \frac{df}{f} \Omega_{\rm GW}^{0}(f) \,.
\end{equation}
The upper limit of this integral $f_{\rm e}$ is the frequency corresponding to the wavelength of a GW equal to the Hubble radius at the end of inflation.  
This energy density redshifts as radiation and, hence, contributes to the total number of relativistic degrees of freedom $N_{\rm eff}$. This contribution as $\Delta N_{\rm eff}$ is given by the expression,
\begin{equation}
\Delta N_{\rm eff} = \frac{8}{7} \left(\frac{11}{4} \right)^{\frac{4}{3}} \frac{\Omega_{\rm GW}^{0}}{\Omega_{\rm \gamma}^{0} }\,.
\end{equation}

\section{Observable predictions}\label{obser}

\subsection{Single field}
In this section, we investigate the effect of a kination-like epoch on leptogenesis and study the observability of gravitational waves in such a non-standard cosmic evolution. To study leptogenesis,
we solve for the Boltzmann equations Eq.~(\ref{asy}) using the estimate of CP asymmetry parameter $\varepsilon$ from Eq.~(\ref{limit}), in the strong washout regime $\Gamma_1/H_1=600$, as pointed in section~\ref{LEP}. We solve the Boltzmann equations
assuming equilibrium RHN abundance and three different choices of $w=0.6,~0.8,~1$. In Tables \ref{t1}-\ref{t3} the benchmark points corresponding to successful leptogenesis scenarios (that produce the observed baryon abundance) are tabulated for the values of $w$ we have considered. Note that, for comparison, we have also tabulated the lepton asymmetry yield ($Y_L^{\rm RD}$) for benchmarks when standard cosmological history (with radiation domination era after inflation) is considered. From the Tables \ref{t1}-\ref{t3}, we observe that the lepton yield in a single kination-like era is significantly enhanced with respect to standard leptogenesis. It is clear from these tables that the benchmarks that could not produce the right baryon asymmetry in standard cosmic history can give rise to the required baryon asymmetry when kination-like epochs are considered.  

\begin{table}[htb]
\centering
 \begin{tabular}{| c | c | c | c | c| c|} 
 \hline
 BM &$\varepsilon$ & $T_{\rm R}/M_1$ & $M_1$ (GeV) & $Y_L$ & $Y_L^{\rm{RD}}$ \\ [0.5ex] 
 \hline
1 & $1.0\times10^{-7}$ & $2.4\times10^{-9}$ & $1.0\times10^{9}$ & $2.69\times10^{-10}$ & $1.13\times10^{-14}$ \\
2 & $5.0\times10^{-7}$ & $2.9\times10^{-8}$ & $5.0\times10^{9}$ &  $2.59\times10^{-10}$ & $5.74\times10^{-14}$ \\
3 & $2.5\times10^{-6}$ & $2.0\times10^{-7}$ & $2.5\times10^{10}$ & $2.59\times10^{-10}$ & $2.68\times10^{-13}$ \\
4 & $1.0\times10^{-5}$ & $1.8\times10^{-6}$ & $1.0\times10^{11}$ & $2.52\times10^{-10}$ & $1.14\times10^{-12}$ \\
 \hline
 \end{tabular}
\caption{\it Benchmark parameters for successful leptogenesis influenced by a single field with e.o.s. $w=0.6$. $Y_L$ denotes the present-day comoving abundance of lepton asymmetry. The respective values of $Y_L^{\rm RD}$ under standard radiation-dominated universe are shown for comparison.}
\label{t1}
\end{table}

\begin{table}[htb]
\centering
 \begin{tabular}{| c | c | c | c | c| c|} 
 \hline
 BM &$\varepsilon$ & $T_{\rm R}/M_1$ & $M_1$ (GeV) & $Y_L$ & $Y_L^{\rm{RD}}$ \\ [0.5ex] 
 \hline
1 & $1.0\times10^{-7}$ & $0.6\times10^{-5}$ & $1.0\times10^{9}$ & $2.65\times10^{-10}$ & $1.13\times10^{-14}$ \\
2 & $5.0\times10^{-7}$ & $2.3\times10^{-5}$ & $5.0\times10^{9}$ &  $2.72\times10^{-10}$ & $5.74\times10^{-14}$ \\
3 & $2.5\times10^{-6}$ & $6.3\times10^{-5}$ & $2.5\times10^{10}$ & $2.64\times10^{-10}$ & $2.68\times10^{-13}$ \\
4 & $1.0\times10^{-5}$ & $2.0\times10^{-4}$ & $1.0\times10^{11}$ & $2.55\times10^{-10}$ & $1.14\times10^{-12}$ \\
 \hline
 \end{tabular}
\caption{\it Same as Table\,\ref{t1} but shown for kination with single scalar field for equation-of-state parameter $w=0.8$.}
\label{t2}
\end{table} 

\begin{table}[htb]
\centering
\begin{tabular}{|c|c|c|c|c|c|}
\hline
BM & $\varepsilon$ & $T_{\rm R}/M_1$ & $M_1~(\mathrm{GeV})$ & $Y_L$ & $Y_L^{\rm RD}$ \\
\hline
1 & $1.0\times10^{-7}$ & $1.4\times10^{-4}$ & $1.0\times10^{9}$  & $2.51\times10^{-10}$ & $1.13\times10^{-14}$ \\
2 & $5.0\times10^{-7}$ & $3.5\times10^{-4}$ & $5.0\times10^{9}$  & $2.47\times10^{-10}$ & $5.74\times10^{-14}$ \\
3 & $2.5\times10^{-6}$ & $6.5\times10^{-4}$ & $2.5\times10^{10}$ & $2.59\times10^{-10}$ & $2.68\times10^{-13}$ \\
4 & $1.0\times10^{-5}$ & $1.3\times10^{-3}$ & $1.0\times10^{11}$ & $2.64\times10^{-10}$ & $1.14\times10^{-12}$ \\
\hline
\end{tabular}
\caption{\it Same as Table\,\ref{t1} but shown for kination with single scalar field for equation-of-state parameter $w=1$.}
\label{t3}
\end{table}

\begin{figure}
\begin{center}
\includegraphics[width=0.49\textwidth]{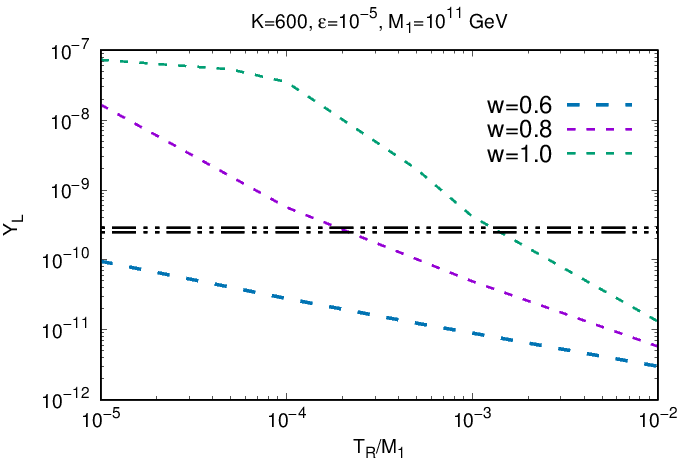}
\includegraphics[width=0.465\textwidth]{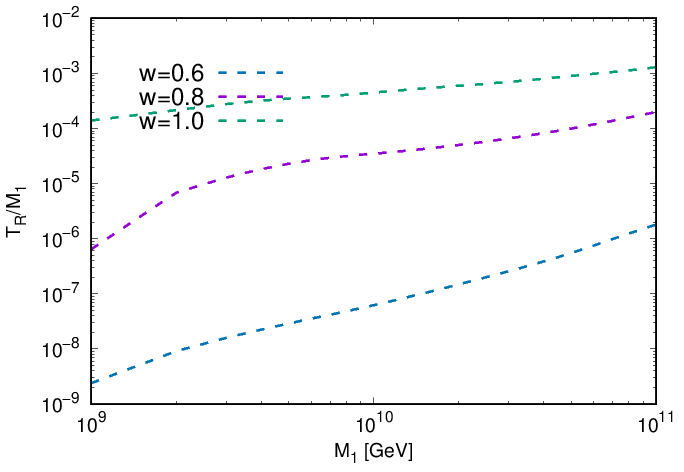}
	\caption{\it Left panel: Variation of present-day abundance of lepton asymmetry $Y_L$ with $T_{\rm R}/M_1$ for fixed RHN mass $M_1$ and asymmetry parameter $\varepsilon$. Blue, purple, and green dashed lines indicate single scalar field driven non-standard cosmological evolution with different e.o.s. $w=0.6,~0.8,~1$. Black horizontal lines show the required abundance of $Y_L$ in agreement with the baryon asymmetry $Y_B$ in the universe. Right Panel: Variation of $T_{\rm R}/M_1$ against the lightest RHN mass $M_1$ is shown for successful leptogenesis in the presence of a single scalar field. Different colors indicate the different choices of $w$. 
    }
\label{fig:figlep1}
\end{center}
\end{figure}

In Fig.~\ref{fig:figlep1} (left panel) we demonstrate the variation of comoving abundance $Y_L$ obtained by solving Eq.~(\ref{asy}) with the ratio $T_{\rm R}/M_1$ for the chosen values of $w$, with a fixed set of parameters $K,~\varepsilon$, and $M_1$,  considering initial equilibrium RHN abundance $Y_{N_1}^{\rm ini}=Y_{N_1}^{\rm eq}$. For all the choices of $w$, we observe comoving abundance $Y_L$
reduces with increased $T_{\rm R}/M_1$. This is obvious as for fixed $M_1$, $T_{\rm R}/M_1$ determines the activity range of the scalar field up to temperature $T_{\rm R}$ scaled with the scale of leptogenesis ($M_1$). A larger $T_{\rm R}/M_1$ indicates that the effect of the scalar field turns off earlier, resulting in larger washout of asymmetry as the universe enters radiation-dominated era. This phenomenon is well established and in agreement with the reporting of earlier studies \cite{Chen:2019etb}, determined by the factor $J_1$ in Eq.~(\ref{factor}), which helps reduce the strong washout set by the decay parameter $K=600$. Since $J_1$ is essentially larger for higher values of $w$, comoving abundance $Y_L$ further enhances for larger choice of $w$, countering the washout of asymmetry. This explains the increased $Y_L$ for $w=1$ when compared with $w=0.8$ (and $w=0.6$) for a fixed ratio $T_{\rm R}/M_1$ depicted in the plots of Fig.~\ref{fig:figlep1}. The horizontal lines denotes the required abundance of $Y_L$ in agreement with the baryon asymmetry $Y_B$ in the universe. The results confirm that the effect of high $w$ weakens the washout of asymmetry in $Y_L$ and generates the correct matter anti-matter asymmetry at a higher $T_{\rm R}/M_1$. On the contrary, for small $w=0.6$, one fails to attain the required lepton asymmetry with $T_{\rm R}/M_1\geq 10^{-5}$. Thus, for smaller $w$, the scalar field may remain active for a large duration to compensate the large washout of asymmetry. This is justified from the Table\,\ref{t1}, showing benchmark values for $w=0.6$ with $T_{\rm R}/M_1$ ranging between $10^{-9}$ to $10^{-6}$.

For further clarification of the effect of the scalar fields with different e.o.s on leptogenesis, in the right panel of Fig.~\ref{fig:figlep1}, we show the variation of the lightest RHN mass $M_1$ with $T_{\rm R}/M_1$ that generates the correct order of matter anti-matter asymmetry for a fixed $K=600$ (following discussion of Eq.~(\ref{decpar})) and the chosen values of $w$. Upon scrutiny, we observe that a larger value of $T_{\rm R}/M_1$ is preferred in order to obtain the desired lepton asymmetry in the presence of a kination-like epoch with a larger $w$. This clearly indicates that the effect of scalar field turns off earlier as high $w$ corresponds to lower washout of the comoving density of $Y_L$. On the other hand, for small $w=0.6$, the scalar field remains active for a long duration, setting the stage for smaller values of $T_{\rm R}$ to generate the required lepton asymmetry. We observe that $T_{\rm R}$ can be as small as a few GeV for $M_1=10^9$ GeV, consistent with the limits from BBN. This finding aligns with the small values of $T_{\rm R}/M_1$ appearing in Table\,\ref{t1} for $w=0.6$ with respect to larger values of $T_{\rm R}/M_1$ for $w=0.8$ and $w=1$ shown in Table\,\ref{t2}-\ref{t3}. In other words, a smaller $w$ indicates larger washout, requiring the scalar field to remain active for a larger duration (smaller $T_{\rm R}/M_1$) to achieve the correct baryon asymmetry.

\begin{figure}
\begin{center}
\includegraphics[width=1\textwidth,height=0.26\textheight]{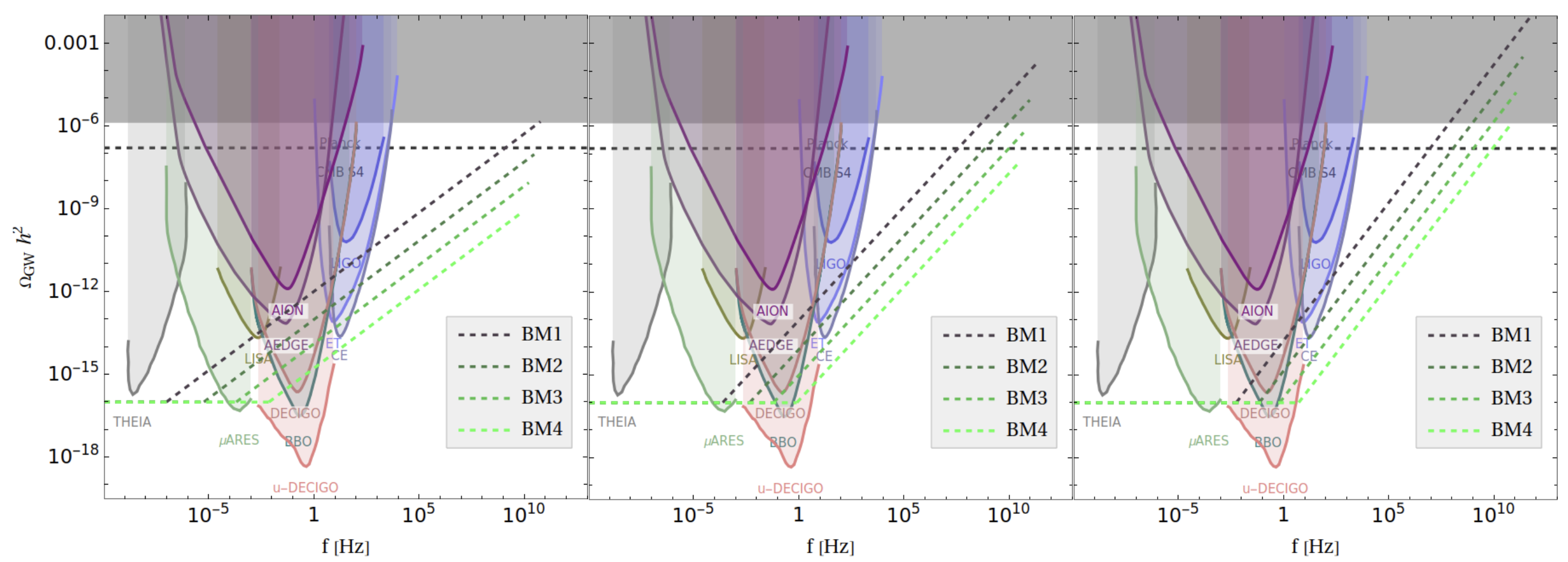}
	\caption{\it In these plots, we have shown the spectral energy density of PGWs for scenarios assuming $w=0.6,~0.8$ and $1$ as mentioned in Table\,\ref{t1}, \ref{t2} and \ref{t3}, respectively. The flat part of the SED corresponds to modes that reenter the Hubble radius during radiation domination. The rise in SED happens for the modes that reenter during a kination-like epoch. It is clear from the plots that higher $w$ corresponds to steeper slope. 
    }
\label{fig:1field}
\end{center}
\end{figure}

\begin{figure}
\begin{center}
\includegraphics[width=0.65\textwidth]{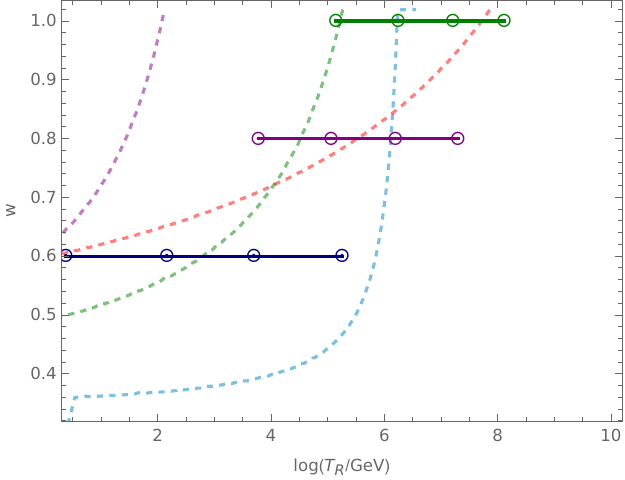}
	\caption{\it In this figure, we have shown the $T_{\rm R}-w$ parameter space for the sigle-field scenario. Solid horizontal lines correspond to the range of $T_{\rm R}$ for different choices of $w$. These are set by the chosen range of RHN mass $M_1=10^{9}-10^{11}$ GeV, that allows for successful leptogenesis, as shown in the right panel of Fig.~\ref{fig:figlep1}. Also shown are the benchmark points tabulated in Table\,\ref{t1}, \ref{t2} and \ref{t3} with deep blue, purple and green $\odot$ markers for the scenarios $w=0.6,~0.8$ and $1$. The purple, green and blue dashed lines correspond to SNR=10 for the future GW observations LISA, ET and DECIGO, respectively. Higher SNR contours are towards the left of the dashed lines shown here, hence regions to the left of the dashed lines have better observability in these experiments. The red dashed line denotes $N_{\rm eff}=0.3$ and the left side of the red dashed line $(N_{\rm eff}>0.3)$ is disallowed from BBN observations.  
	 }
\label{fig:snr1field}
\end{center}
\end{figure}

We now discuss the effect of a single kination-like epoch on the SED of PGWs. In Fig. \ref{fig:1field} we have plotted the SED of PGWs for such a scenario, along with sensitivity curves taken from  LISA~\cite{LISA:2017pwj, Baker:2019nia}, BBO~\cite{Crowder:2005nr,Corbin:2005ny,Harry:2006fi}, DECIGO~\cite{Seto:2001qf,Kawamura:2006up,Yagi:2011wg}, CE~\cite{LIGOScientific:2016wof,Reitze:2019iox}, ET~\cite{Punturo:2010zz, Hild:2010id,Sathyaprakash:2012jk, Maggiore:2019uih}, $\mu-$ARES~\cite{Sesana:2019vho}, LVK~\cite{Harry:2010zz,LIGOScientific:2014pky,VIRGO:2014yos,LIGOScientific:2019lzm}. To quantify the observability of PGW in these detectors, signal-to-noise ratio (SNR) ~\cite{Thrane:2013oya,Caprini:2015zlo} is defined,
 \begin{align}
 	\text{SNR}_\text{exp} \equiv \left\{ 2 t_\text{obs} \int_{f_\text{min}}^{f_\text{max}} d f \left[ \frac{h^2 \Omega_\text{GW}(f)}{h^2\Omega_\text{eff}(f)}\right]^2\right\}^{1/2}\,,
 \end{align}
 where $t_\text{obs}$ is the observation time and $\Omega_\text{eff}(f)$ denotes to the noise curve of the GW observation operating between the frequency range $f_\text{min}$ to $f_\text{max}$.

In the three panels of Fig.\,\ref{fig:1field} we have shown the SED of PGWs for the benchmark points mentioned in Tab.\,\ref{t1}, \ref{t2} and \ref{t3}, respectively. 
As expected from the discussion of section\,\ref{sec:GWspec}, the slope of SED of PGWs in the rightmost panel of Fig.\,\ref{fig:1field} (for $w=1$) is steeper than in the scenarios with $w=0.8$ or $w=0.6$. 
The same benchmark points are shown in Fig.\,\ref{fig:snr1field} by the blue, purple and green $\odot$ markers. In Fig.\,\ref{fig:snr1field}, the purple, green and blue dashed lines correspond to SNR $ =10$ for the future GW observations LISA, ET and DECIGO, respectively, while the value of SNR increases towards the left. The left side of the red dashed line $(N_{\rm eff}>0.3)$ is disallowed. In the left panel of  Fig.\,\ref{fig:1field}, we observe that BM1 touches the region ruled out by $N_{\rm eff}$ bound of Planck 18 \cite{Akrami:2018odb}, denoted by the gray shaded region in the top. BM1 and BM2 cross the sensitivity curve of ET, whereas all four benchmark points cross the sensitivity curve of DECIGO. 
These facts can also be clearly seen from the blue $\odot$ markers and the SNR $ =10$ curves in Fig.\,\ref{fig:snr1field}. 
In the middle panel of the same figure, we see that BM1 and BM2 cross the region ruled out by the $N_{\rm eff}$ bound. These are the same two benchmark points which cross the sensitivity curve of DECIGO. None of the four benchmark points in Table\,\ref{t2} cross the LISA sensitivity curve, whereas only BM1 crosses sensitivity curve of ET. These facts are also evident from the purple $\odot$ markers and the SNR $ =10$ curves in Fig.\,\ref{fig:snr1field}.   
Similarly, for the scenario with $w=1$, we see that BM1, BM2 and BM3 of Table\,\ref{t3} cross the region ruled out by the $N_{\rm eff}$ bound, while BM4 is safe from it. However, only BM1 among these benchmarks crosses the DECIGO and ET sensitivity curves. So, none of the benchmark points are simultaneously safe from $N_{\rm eff}$ bound and observable in the GW observatories. 
In summary, Fig.\,\ref{fig:snr1field} shows the fact that, for $w=0.6$ all the benchmark points are observable in DECIGO and all points for this scenario are safe from the $N_{\rm eff}$ bound. On the other hand, none of the benchmark points in Table\,\ref{t2} and \ref{t3} can be observed in the aforementioned GW experiments while also being safe from the $N_{\rm eff}$ bound of Planck 18 simultaneously. However, the purple straight line connecting the benchmark points for $w=0.8$ suggests that some range of $M_1$ interior to $5\times10^9-2.5\times10^{10}$ GeV is safe from the $N_{\rm eff}$ bound, while being observable in DECIGO.

\subsection{Multiple fields}

\begin{table}[htb]
\centering
\begin{tabular}{| c | c | c | c | c | c |}
\hline
BM & $x=T_2/T_{\rm R}$ & $\varepsilon$ & $T_{\rm R}/M_1$ & $M_1$ (GeV) & $Y_L$ \\ [0.5ex]
\hline

1 &  & $1.0\times10^{-7}$ & $3.0\times10^{-4}$ & $1.0\times10^{9}$  & $2.85\times10^{-10}$ \\
2 & 10 & $1.0\times10^{-6}$ & $6.9\times10^{-4}$ & $1.0\times10^{10}$ & $2.65\times10^{-10}$ \\
3 &  & $1.0\times10^{-5}$ & $1.25\times10^{-3}$ & $1.0\times10^{11}$ & $2.60\times10^{-10}$ \\
\hline

1 &  & $1.0\times10^{-7}$ & $6.7\times10^{-5}$ & $1.0\times10^{9}$  & $2.69\times10^{-10}$ \\
2 & 100 & $1.0\times10^{-6}$ & $1.47\times10^{-4}$ & $1.0\times10^{10}$ & $2.81\times10^{-10}$ \\
3 &  & $1.0\times10^{-5}$ & $2.7\times10^{-4}$ & $1.0\times10^{11}$ & $2.70\times10^{-10}$ \\
\hline

1 &  & $1.0\times10^{-7}$ & $3.1\times10^{-6}$ & $1.0\times10^{9}$  & $2.77\times10^{-10}$ \\
2 & $10^4$ & $1.0\times10^{-6}$ & $7.4\times10^{-6}$ & $1.0\times10^{10}$ & $2.72\times10^{-10}$ \\
3 &  & $1.0\times10^{-5}$ & $1.9\times10^{-5}$ & $1.0\times10^{11}$ & $2.84\times10^{-10}$ \\
\hline

1 &  & $1.0\times10^{-7}$ & $1.9\times10^{-7}$ & $1.0\times10^{9}$  & $2.73\times10^{-10}$ \\
2 & $10^6$ & $1.0\times10^{-6}$ & $1.2\times10^{-6}$ & $1.0\times10^{10}$ & $2.74\times10^{-10}$ \\
3 &  & $1.0\times10^{-5}$ & $1.4\times10^{-5}$ & $1.0\times10^{11}$ & $2.82\times10^{-10}$ \\
\hline

\end{tabular}
\caption{\it Benchmark parameters for successful leptogenesis in the presence of multiple scalars with e.o.s. changing from $w_2=1$ to $w_1=2/3$. Transition from the $\phi_2$-dominated to the $\phi_1$-dominated Universe is characterized by $T_2$, while the transition to radiation domination occurs at the temperature of $T_{\rm R}$.}
\label{t4}
\end{table}

\begin{figure}
\begin{center}
\includegraphics[width=0.495\textwidth]{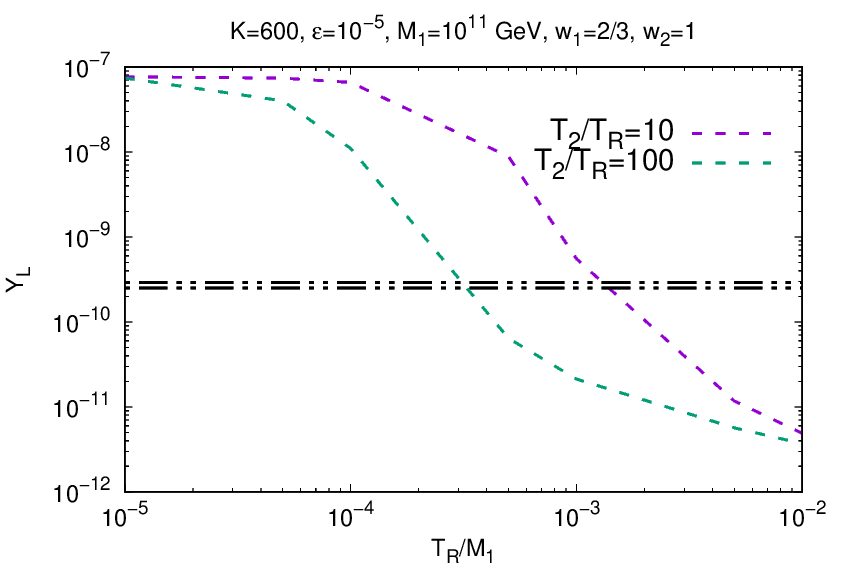}
\includegraphics[width=0.495\textwidth]{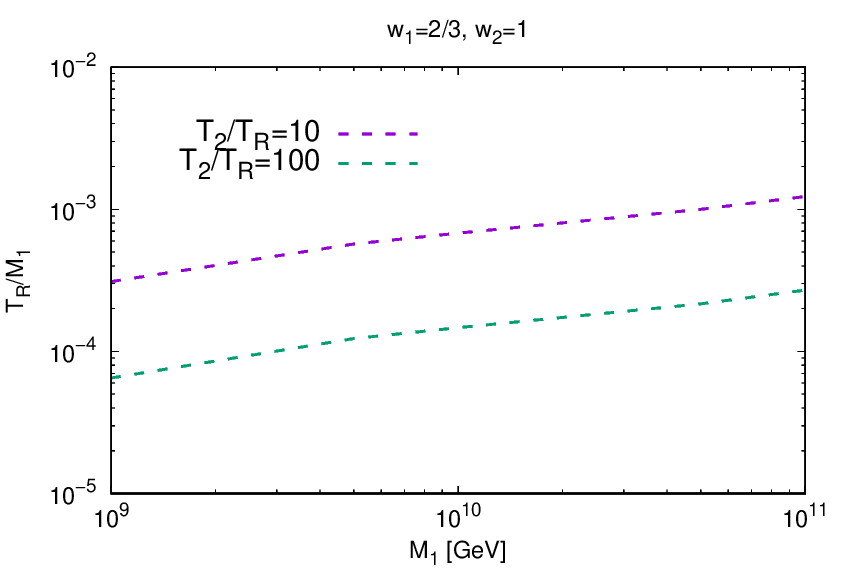}
	\caption{\it Left panel: Variation of lepton asymmetry $Y_L$ against $T_{\rm R}/M_1$ shown for leptogenesis assisted by multiple scalars with different e.o.s. ($w_2=1,~w_1=2/3$) for fixed RHN mass $M_1$ and asymmetry parameter $\varepsilon$. Purple (green) dashed line indicates the duration $T_2/T_{\rm R}=10$ ($T_2/T_{\rm R}=100$) of the $\phi_1$-dominated epoch, while $T_{\rm R}/M_1$ sets the start of radiation domination. Black horizontal lines depict the range of $Y_L$ in agreement with matter antimatter asymmetry of the universe. Right Panel: Variation of $T_{\rm R}/M_1$ against lightest RHN mass $M_1$ shown for successful leptogenesis under the influence of multiple scalar fields with colors indicating the choice of $x=T_2/T_{\rm R}$.}
\label{fig:figlep2}
\end{center}
\end{figure}

We now discuss leptogenesis in the scenario where two scalar fields dominate the energy density of the universe sequentially before radiation domination.
Table~\ref{t4} presents some benchmark points where the effects of two scalar fields are considered, resulting in back-to-back kination-like epochs with $w_2=1$ and $w_1=2/3$. These results are obtained from the solutions to the Boltzmann equations for leptogenesis with multiple scalar fields Eq.~(\ref{asy2}). In this present scenario, radiation domination occurs after two-step kination. 
The energy density of $\phi_2$ field dominates at temperature $T>T_2$ resulting an e.o.s of $w_2=1$. 
Subsequently, for temperature $T_2>T>T_{\rm R}$, the expansion of universe is governed by the $\phi_1$ field, giving rise to an e.o.s of $w_1=2/3$. 
Therefore, $T_2$ and $T_{\rm R}$ specify the transitions from domination of $\phi_2$ to $\phi_1$ and $\phi_1$ to radiation, respectively. 
As mentioned before in section~\ref{LEP}, the two new scales determining the period of domination of the fields are set by the dimensionless parameters $x=T_2/T_{\rm R}$ and $T_{\rm R}/M_1$. For this setup with multiple fields, in Fig.~\ref{fig:figlep2}, we show the variation of $Y_L$ with $T_{\rm R}/M_1$  (with fixed $\varepsilon=10^{-5},~K=600$), and required $T_{\rm R}/M_1$ for change in $M_1$ to obtain the correct baryon asymmetry in presence of two scalar fields with two choices of $T_2/T_{\rm R}$.
Comparison of the plots of Fig.~\ref{fig:figlep1} with Fig.~\ref{fig:figlep2} reveals that the change in the parameter $T_2/T_{\rm R}$ in multiple-field scenario produces similar effects as changing $w$ for the single-field case. 
It is observed from Fig.~\ref{fig:figlep2} (left panel) that at very low $T_{\rm R}/M_1\simeq 10^{-5}$ and very high $T_{\rm R}/M_1\simeq 10^{-2}$ the curves with different $T_2/T_{\rm R}$ almost merge together. 
A small $T_{\rm R}/M_1$ indicates the field $\phi_1$ is active for a long duration that nullifies the washout effect of the asymmetry significantly. 
For weak washout, $Y_L\propto \varepsilon$ as the washout contribution in Eq.~(\ref{asy2}) is absent. Therefore, for complete absence of the washout, $Y_L$ achieves a maximum value depending on the asymmetry parameter $\varepsilon$ only. This justifies the nature of the plots for small $T_{\rm R}/M_1$ saturating at a specific value.

\begin{figure}
\begin{center}
\includegraphics[width=0.49\textwidth]{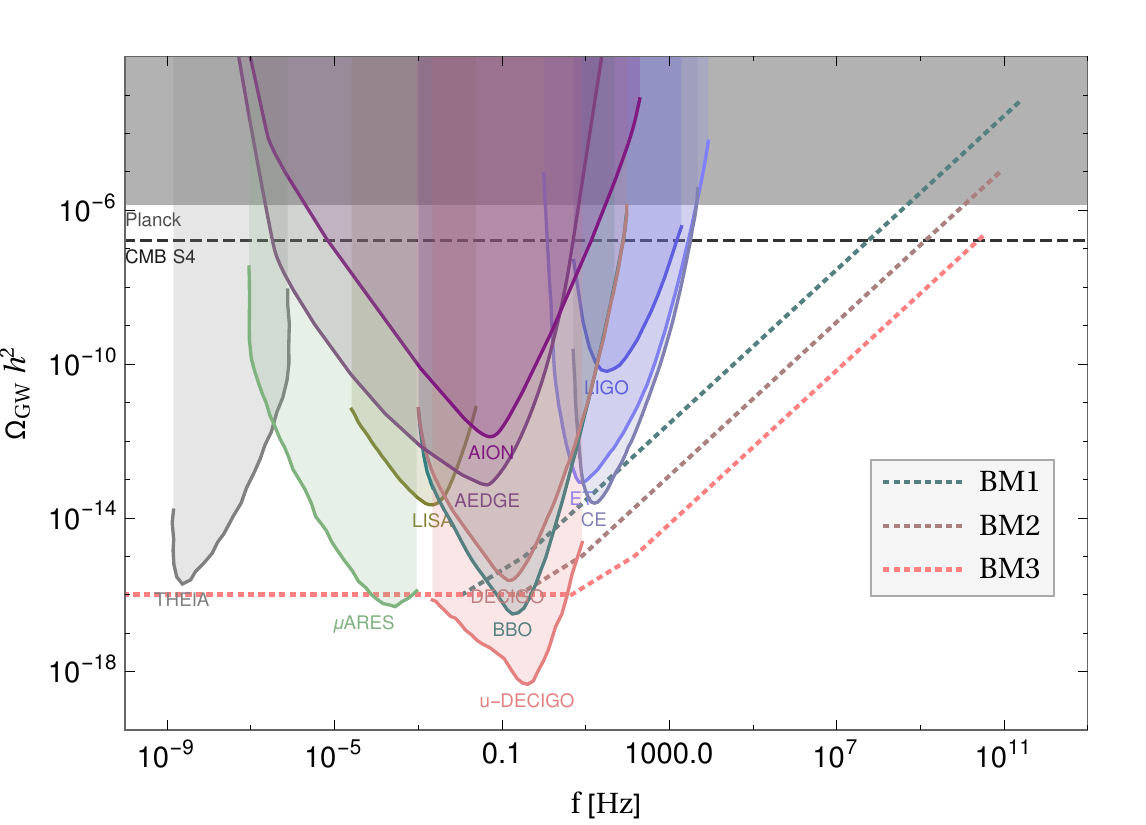}
\includegraphics[width=0.49\textwidth]{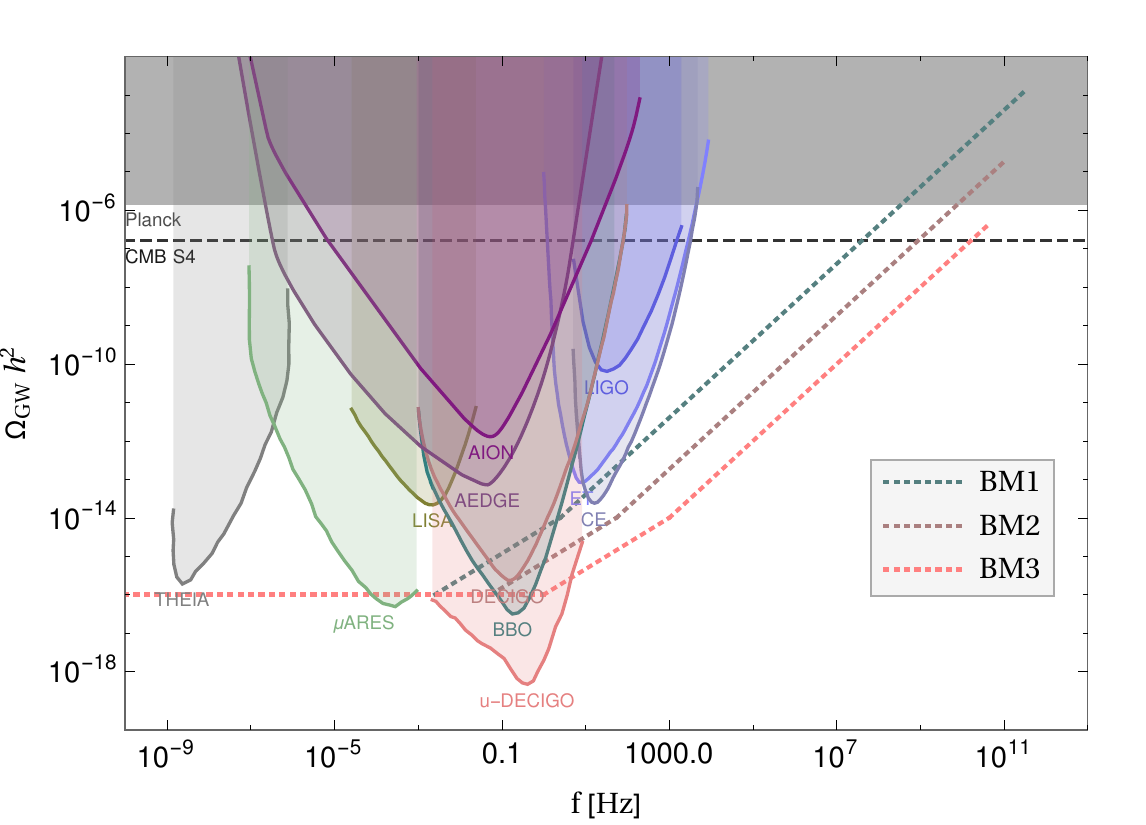}
	\caption{\it In this plot, we have shown the spectral energy density of PGWs for scenarios with $x=10,~100$, assuming $w_2=1$ and $w_1=2/3$ as mentioned in Table \ref{t4}. The flat part of the SED corresponds to modes that reenter the Hubble radius during radiation domination. The rise in SED happens for the modes that reenter during $w_1$ or $w_2$ domination. The changes in slope of the SED, corresponding to changes in the equation of state, are clearly visible in the SED.}
\label{fig:2field}
\end{center}
\end{figure}

In a similar fashion, at high $T_{\rm R}/M_1$, the radiation domination effects start earlier and thus reduce the lepton asymmetry to the same extent for both scenarios, independent of $T_2/T_{\rm R}$, almost nullifying the effects of the scalar fields. However, in the regime $10^{-5}\leq T_2/T_{\rm R} \leq 10^{-2}$, the effects are distinguishable for different $T_2/T_{\rm R}$. With an increased value of $T_2/T_{\rm R}$, a fixed $T_{\rm R}/M_1$ reduces the lepton asymmetry as the effect of the second scalar field ceases earlier. 
Consequently, to generate the desired abundance of lepton asymmetry with larger $T_2/T_{\rm R}$, one requires to reduce the value of $T_{\rm R}/M_1$ for a fixed $M_1$. This effect is clearly visible in the right panel of Fig.~\ref{fig:figlep2}, and further validated with the findings in Table \,\ref{t4}.

\begin{figure}
\begin{center}
\includegraphics[width=0.65\textwidth]{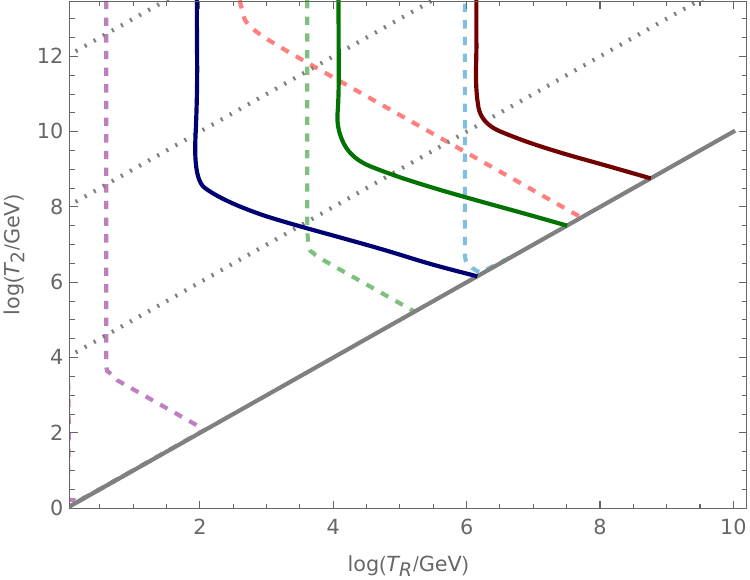}
	\caption{\it We have shown the regions in the $T_{\rm R}-T_2$ parameter space where right relic lepton asymmetry can be achieved, as well as the regions observable in future GW observations. The dark blue, green and brown lines correspond to the correct yield of lepton asymmetry for masses $M_1=10^9, 10^{10}, 10^{11}$ GeV, respectively. The purple, green and blue dashed lines correspond to SNR $= 10$ for the future GW observations LISA, ET and DECIGO, respectively. Higher SNR contours are towards the left of the dashed lines shown here; hence, to the left of the dashed lines have better observability in these observations. The red dashed line denotes $N_{\rm eff}=0.3$, and the left side of the red dashed line $(N_{\rm eff}>0.3)$ is disallowed from BBN observations. 
So, the only possible region allowed by successful leptogenesis, allowed by $N_{\rm eff}$ bounds and observable in future experiments (DECIGO only) is the small triangular area surrounded by the red and blue dashed lines ($N_{\rm eff}$ and DECIGO). A smaller triangular region surrounded by the red and green dashed line is observable in ET, DECIGO and is allowed by successful leptogenesis and $N_{\rm eff}$ bounds. For this plot we have assumed $w_2=1$ and $w_1=2/3$. }
\label{fig:snr2field}
\end{center}
\end{figure}

From the comparison of Fig.~\ref{fig:figlep1} with Fig.~\ref{fig:figlep2}, it 
can easily be concluded that the effect of single or multiple scalar field, although present prominently in the context of leptogenesis, are almost similar. 
This is due to that effect of single or multiple scalars appear in the Boltzmann's equations of leptogenesis through $J_1$ or $J_2$. Therefore, for a certain choice of $\varepsilon$, and $M_1$, with $J_1\simeq J_2$, the effect on the lepton asymmetry is indistinguishable, although they appear through different mechanisms. It is also observed from the plots of Fig.~\ref{fig:figlep1} and Fig.~\ref{fig:figlep2} that the effect of change in $w$ in the case of single scalar field is more or less equivalent to changes in $T_2/T_R$ with fixed values of $w_2,~w_1$ when multi-step kination-like effects are considered. However, this apparent indistinguishable nature of scalar fields can be resolved via observing their effect in SED of PGWs.

In the left and right panel of Fig.\,\ref{fig:2field} we show the SED of PGW for the first two parts of Table\,\ref{t4}, \textit{i.e.} for $x=T_2/T_{\rm R}=10$ and $100$ respectively. In these plots it is easy to see that the slope of the SED changes from being flat for the modes entering during radiation domination to some positive slope of the modes entering during the kination-like epoch with e.o.s $w_1$ and eventually becomes steeper for the modes entering during the epoch with e.o.s $w_2>w_1$. It is also evident from the left ($x=10$) and the right ($x=100$) panel that increasing the value of $x$, \textit{i.e.} having a longer epoch with $w_1$, increases the range of frequencies with SED slope of $2(3w_1-1)/(1+3w_1)$.
In Fig.\,\ref{fig:snr2field} we show the $T_{\rm R}-T_2$ parameter space. The dark blue, green, and brown curves correspond to the right yield of lepton asymmetry for mass $M_1=10^9, 10^{10}, 10^{11}$ GeV, respectively. We show the SNR = 10 curves for the future GW observations LISA, ET and DECIGO with purple, green and blue dashed lines, respectively, while the SNR increases towards the left of the curves. The red dashed line denotes $N_{\rm eff}=0.3$, the left side of which $(N_{\rm eff}>0.3)$ is disallowed. So, from Fig.\,\ref{fig:snr2field}, we can conclude that for $M_1=10^{11}$ GeV, the whole curve is allowed by the $N_{\rm eff}$ bound, but not likely observable in LISA, ET and DECIGO. On the other hand, the curve denoting $M_1=10^{9}$ GeV is observable in DECIGO fully and partially in ET; however, the whole curve is disallowed by the $N_{\rm eff}$ bound. For the curve $M_1=10^{10}$ GeV, a small part near the top (vertical part of the solid green curve) is allowed by the  $N_{\rm eff}$ bound while being observable in DECIGO. The small triangular region surrounded by the red and blue dashed curves is observable in DECIGO and is allowed by successful leptogenesis and $N_{\rm eff}$ bounds. 

\medskip

\section{Discussions and Conclusions}\label{disc}
In this work, we investigate the possible early universe signature of leptogenesis as the origin of matter anti-matter asymmetry in the universe. The simplest route to the matter anti-matter asymmetry via leptogenesis can be orchestrated via inclusion of heavy right handed neutrinos. The CP-violating decay of the lightest RHN not only provides the origin of baryon asymmetry, but also acts as the source for generation of light neutrino mass via the type-I seesaw mechanism. However, this conventional leptogenesis with heavy RHNs faces serious drawbacks.
Although a sufficient amount of asymmetry can be produced initially, for standard cosmic history, it faces a strong washout effect for a lightest RHN mass of $M_1\approx 10^9-10^{11}$ GeV, and the asymmetry depletes significantly failing to maintain the observed present-day baryon abundance. 
In addition, the flavor effects come into play, adding further difficulties to the process. 
Another challenge is the non-verifiability of such phenomena at terrestrial experiments, as the process of leptogenesis occurs at a very high scale. 

In order to circumvent the shortcomings of leptogenesis through heavy RHN, we consider an alternate cosmological history with one or multiple kination-like epochs (due to domination of scalar fields).
The presence of such scalar fields in early universe is well motivated from four-dimensional superstring models associated with D-branes. 
These scalar fields, resulting in kination-like epochs in the early universe with different e.o.s. ($w$), have a threefold effect on leptogenesis. 
Firstly, they alter the expansion rate of the universe, faster than radiation domination, and nullify the strong washout of the lepton asymmetry. 
Secondly, the faster expansion in presence of scalars delays the flavor effects and broadens the scope of vanilla leptogenesis. 
Therefore, the combined effect allows the net lepton asymmetry to survive and explain the observed baryon asymmetry in the universe. 
Finally, such kination-like epochs leave their imprints on the SED of PGWs, thus leaving a detectable signature of early universe evolution at high temperature. 
We consider two different scenarios where the evolution history of the universe is changed by the influence of one single scalar field and multiple scalar fields for our analysis and summarize the findings of our investigation in the following points:
\begin{itemize}
   \item We find that Brane-modified cosmology with both single or multiple fields can assist leptogenesis with heavy RHNs to generate the observed amount of matter anti-matter asymmetry, while the standard radiation-dominated leptogenesis fails to achieve the same due to strong washout effects and hence dilution of asymmetry. In addition to the enhanced expansion rate of universe, the scalar fields also mitigate the flavor effects in leptogenesis, enhancing the opportunity for vanilla leptogenesis.

   \item For single kination-like epoch, we observe that the baryon asymmetry can be achieved for different value of parameters $w$ and $T_{\rm R}$ where $w$ denotes the e.o.s. and $T_{\rm R}$ signifies the temperature during transition into radiation-dominated universe (see Table\,\ref{t1}-\ref{t3}). Similarly, from Table\,\ref{t4}, we observe that the same can occur in the presence of domination of multiple scalar fields with new parameters $w_1,~w_2,~T_2,~T_{\rm R}$. However, it is difficult to speculate whether leptogenesis is influenced by a single scalar or multiple scalars from baryon asymmetry. Conversely, presence of one or multiply kination-like epochs can be realized through observation of the modified SED of PGWs. While a single scalar field generates a single positive slope in the otherwise scale invariant SED, for multiple scalar fields one expects a SED containing multiple slopes corresponding to the different e.o.s of the early universe (see Fig.~\ref{fig:1field},~Fig.\ref{fig:2field}). Therefore, through observation of SED of PGWs, we can probe the evolution of the early universe and distinguish the effects of scalars on the evolution of the universe and generation of baryon asymmetry.

    \item Detailed analysis of the modified expanding universe with the influence of one scalar field ($w=1$) reveals that vanilla leptogenesis is constrained severely by the $N_{\rm eff}$ bounds (see Fig.~\ref{fig:snr1field}). For a kination-like effect with $w=0.8$, we find a small parameter space which is observable in DECIGO while being allowed by the $N_{\rm eff}$ bounds. We find that a smaller $w=0.6$ is preferred in terms of observability, i.e. the complete parameter space for successful leptogenesis ($M_1=10^9-10^{11}$ GeV) is allowed by the $N_{\rm eff}$ bounds and detectable by DECIGO and partly by ET.

    \item Study of SNR to observationally probe the effect of successful leptogenesis under the influence of multiple scalar-driven evolution of the universe provides some interesting results (see Fig.~\ref{fig:snr2field}). While a pure kination $w=1$ is ruled out when only one scalar field is considered, in the presence of multiple scalars, kination effect ($w_2=1$) is allowed when assisted by another scalar with $~w_1=2/3$. However, constraints from BBN rule out leptogenesis for the lightest RHN mass $M_1=10^{9}$ GeV, and only a small region of parameter space survives for $M_1=10^{10}$ GeV, which is detectable by DECIGO. Although the parameter space for $M_1=10^{11}$ GeV survives the $N_{\rm eff}$ bounds, it has very low detection prospects and may only be traced with future detectors with better sensitivity. 

\end{itemize}

Therefore, Brane-inspired non-standard cosmology not only elucidates high scale leptogenesis successfully, relaxing strong washout and flavor effects, but also enhances the detection prospects through the observation of PGWs. The non-standard cosmology is driven by the presence of multiple non-interacting scalars with different e.o.s. altering the expansion history of Universe. Although it allows successful leptogenesis under the influence of single or multiple scalars, the study of PGWs, along with the bounds from $N_{\rm eff}$, constrains the lightest RHN mass ranging between $M_1\simeq 10^9-10^{11}$ GeV, depending on the e.o.s. parameter $w$ for single field scenario and $w_1,~w_2$ for multiple-field scenario. Interestingly, SED of PGWs allows us to discriminate the effects of single or multiple scalar driven expansion of the universe and hence acts as a indirect probe of high scale leptogenesis in modified cosmology. Future observations in the light of PGWs are expected to further shift the lightest RHN mass above $M_1\geq 10^{11}$ GeV, allowing a prominent probe of high scale leptogenesis.

\section*{Acknowledgement}

The authors acknowledge Anish Ghoshal for discussions during initial stage of the project. ADB acknowledges financial support from DST, India, under grant number IFA20-PH250 (INSPIRE Faculty Award). AP thanks the Indo-French Centre for the Promotion of Advanced Research for supporting the postdoctoral fellowship through the proposal 6704-4 under the Collaborative Scientific Research Programme and Anusandhan National Research Foundation for the National PostDoctoral fellowship, through file 
number: PDF/2025/002087.

\bibliographystyle{JHEP}

\bibliography{ref}

\end{document}